\documentclass[12pt,preprint]{aastex}

\def\lea{\mathrel{<\kern-1.0em\lower0.9ex\hbox{$\sim$}}}
\def\gea{\mathrel{>\kern-1.0em\lower0.9ex\hbox{$\sim$}}}

\slugcomment{Submitted to The Astrophysical Journal}

\shorttitle{Willman 1}
\shortauthors{Siegel et al.}

\begin{document}

\title{Trimming Down the Willman 1 dSph\thanks{
Based on observations obtained with the Hobby-Eberly Telescope, which is a joint project of the 
University of Texas at Austin, the Pennsylvania State University, Stanford University, Ludwig-Maximilians-
Universit\"{a}t M\"{u}nchen, and Georg-August-Universit\"{a}t G\"{o}ttingen.
}}

\author{Michael H. Siegel and Matthew D. Shetrone}

\affil{University of Texas, McDonald Observatory, 1 University Station, C1402, Austin TX, 78712
(siegel@astro.as.utexas.edu, shetrone@astro.as.utexas.edu)}

\author{Michael Irwin}

\affil{Institute of Astronomy, Madingley Road, Cambridge, CB3 OHA, UK (mike@astr.cam.ac.uk)}

\begin{abstract}
Willman 1 is a small low surface-brightness object identified in the Sloan Digital Sky Survey
and tentatively
classified as a very low luminosity dSph galaxy. Further study has supported this classification
while hinting that it may be undergoing disruption by the Milky Way potential.
In an effort to better constrain the nature of Willman 1, we present
a comprehensive analysis of the brightest stars in a 0.6 square degree field centered
on the overdensity.
High-resolution HET spectra of two previously identified Willman 1 RGB stars show that one is
a metal-rich foreground dwarf while the other is a metal-poor giant.  The one RGB star that
we confirm as a member of Willman 1 has a low metallicity ([Fe/H]=-2.2) and
a surprisingly low $\alpha$-element abundance ([$\alpha/Fe$]=-0.11).
Washington+$DDO51$ photometry indicates that 2-5
of the seven brightest Willman 1 stars identified in previous studies are actually dwarf stars, 
including some of the more metal-rich stars that have been used
to argue both for an abundance spread and a more metal-rich stellar population than galaxies
of similar luminosity.  The remaining stars are too blue or too faint for
photometric classification.  The Washington+$DDO51$ photometry identifies three potential RGB stars in
the field but HET spectra show that they are background halo stars.
Time series photometry identifies
one apparent variable star in the field, but it is unlikely to be associated with Willman 1. 
Our wide-field survey indicates that over 0.6 square degrees, Willman 1 does not have a single RR Lyrae
star, a single BHB star or a single RGB star beyond its tidal radius.
While our results confirm that Willman 1 is most likely a low-luminosity metal-poor dSph galaxy, the
possibility remains that it is a tidally disrupted metal-poor globular cluster.
\end{abstract}

\keywords{galaxies: dwarf; galaxies: individual (Willman 1); galaxies: abundances; galaxies: photometry; Galaxy: halo}

\section{Introduction}

Willman 1 is a small diffuse low-luminosity object identified by Willman et al. (2005, hereafter W05) from an 
analysis of the
Sloan Digital Sky Survey. Its size and luminosity were found to be consistent with both
low-luminosity globular clusters  
(e.g., Palomar 1, E3, AM 4, BH 176, Sau A and AM 1) and low-luminosity
dSph galaxies (see Figure 10 of W05).  Further investigation revealed
irregular structure, possible
mass segregation and a spatial size too large for Willman 1 to be bound without substantial amounts
of dark matter
(Willman et al. 2007, hereafter W07). Spectroscopic study (Martin et al. 2007; hereafter M07)
showed that Willman 1's red giant branch (RGB) stars 
appear to form a velocity group at -12.3 km s$^{-1}$ with a moderate
velocity dispersion (4.3 km s$^{-1}$). The latter implies a high (470-700)
mass to light (M/L) ratio that would potentially be enough to bind Willman 1 at its 
present galactocentric distance.
Additionally, M07 show an apparent spread in abundance with a mean metallicity 
of [Fe/H]=-1.5 -- the latter being consistent with the abundance
estimate from main sequence photometry (W07; M07). The combination of abundance spread and
high dark matter content would indicate that Willman 1 is a dSph galaxy -- the
lowest luminosity dSph known to be orbiting the Galaxy.

However, if Willman 1 is a dSph galaxy, its properties depart significantly from the established 
patterns of
dSph properties (Figure 1). The M/L ratio and [Fe/H] abundance of dSph galaxies have been 
shown to follow total luminosity (see, e.g., Mateo 1998), a trend used to argue
that the dSph galaxies are of similar mass and origin.
M07 noted and we confirm that Willman 1's apparent M/L ratio is an order of magnitude 
lower than the linear trend established in Milky Way dSph galaxies -- although it is similar
to other dSph galaxies of similar luminosity.
Willman 1 is also significantly more 
metal-rich that dSph galaxies of comparable luminosity.

Simon \& Geha (2007, hereafter SG07) used similar departures from the established $M_V-[Fe/H]$
and $M_V-M/L$ trends to argue that the Coma and Ursa Major II dSph galaxies could be the tidally
disrupted remnants of larger forebears. However, while Ursa Major II appears to overlap the ``orphan stream"
(Belokurov et al. 2007), neither Coma nor Willman 1 are known to overlap a halo stream.
This is particularly problematic in the case of Willman 1. If Willman 1's relatively 
high metallicity
reflects its initial, rather than current, luminosity, Figure 1 would indicate that
Willman 1's tidal stream would contain {\it several thousand} times more mass than its core.

\begin{figure}[h]
\epsscale{1.0}
\plotone{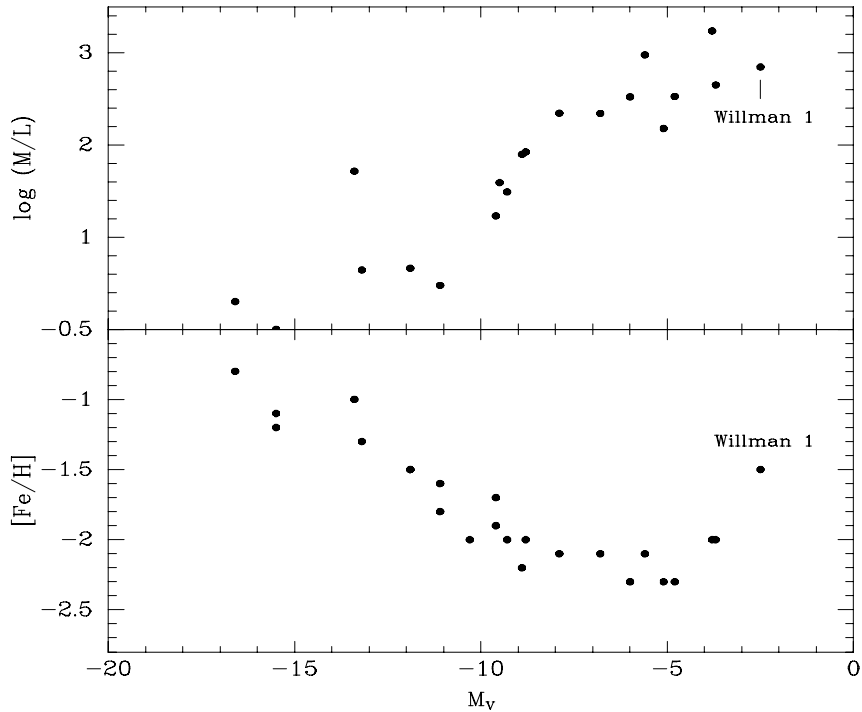}
\caption{Correlations of dSph M/L ratio and [Fe/H] to $M_V$. Data are taken from Mateo et al. (1998),
Munoz et al. (2006), Zucker et al. (2006), SG07 and M07.}
\end{figure}

In this paper, we add several pieces of information to the previous studies of Willman 1 in an effort
to clarify its nature.
We improve membership analysis and provide better constraint on the abundance distribution
using optimized spectra of two supposed Willman 1 members and Washington+$DDO51$ 
photometry (\S3.1). We also survey Willman 1 for RR Lyrae stars (\S3.2) and use wide-field
photometry to probe the extended distribution of the putative dwarf (\S3.3). 

\section{Observations and Data Reduction}

\subsection{HET Spectroscopy}

We obtained spectra of the two brightest Willman 1 RGB stars identified by M07 and three bright
RGB candidates
identified from Washington photometry (\S3.1) using the Hobby-Eberly Telescope (HET) and
High Resolution Spectrograph (HRS, Tull 1998). Data were obtained as part of
standard queue scheduled observing (Shetrone et al. 2007) in November 2006 and
April 2007.
HRS was configured to achieve R=18,000 over the spectral range of
4840-6820\AA\ with the 600g5822 cross-disperser. The November data, which
were used for abundance analysis, consist of multiple 2250-2700 second exposures taken
over several nights to achieve total exposure times of 4500-16200 seconds and
SNR of 23-30 per resolution element.
The April data, taken to determine radial
velocities, have total exposure times between 600 and 720 seconds to achieve SNR $>$ 15 per resolution
element.

The spectra were reduced with IRAF's\footnote{IRAF (Image Reduction and
Analysis Facility) is distributed by the National Optical Astronomy
Observatories, which are operated by the Association of Universities
for Research in Astronomy, Inc., under contract with the National
Science Foundation.} ECHELLE and TWODSPEC scripts.  The standard IRAF scripts 
for overscan removal, bias subtraction, flat fielding and scattered light
removal were employed.  The sky fiber spectra were extracted separately 
and the two resulting sky spectra were combined and subtracted from
the target spectra. We confirmed that no residual solar spectrum remained
in the target spectra by cross-correlating to the solar
atlas of Hinkle et al. (2000) and looking for a peak at zero relative velocity.
The target spectra were shifted to the heliocentric rest frame by 
cross-correlating with the Hinkle et al. (2000) Arcturus atlas and combined in the rest frame.

\subsection{MDO 0.8m Photometry}

We observed Willman 1 with the McDonald Observatory 0.8m telescope and Prime Focus Corrector (PFC)
on UT 19-20 March 2007 and 16-19 April 2007.  Photometry was obtained in the broadband $BV$,
Washington $MT_2$ and intermediate-band $DDO51$ filters. Although unguided, the 0.8m telescope is stable
enough over 600 second time intervals to produce excellent wide-field photometry at the coarse
(1.3 "/pixel) pixel
scale of the PFC with little image ellipticity.  In an effort to identify any variable stars in the field, we
observed ten $B$-band epochs over the course of the observing runs. Total observing times were
6000, 600, 600, 1200 and 6000 seconds in the $B$, $V$, $M$, $T_2$ and $DDO51$ filters, 
respectively.
The 46\farcm0 field
of the PFC should easily enclose the entirety of the Willman 1 object, which has a measured
half-light radius of $r_h=1\farcm9$ and a limiting radius of 
$\sim10\farcm0$ (W05, W07 and M07). The PFC covers nearly twice the area of 
sky as W07's and M07's photometric
surveys, although to much shallower depth.

\begin{figure}[h]
\epsscale{1.0}
\plotone{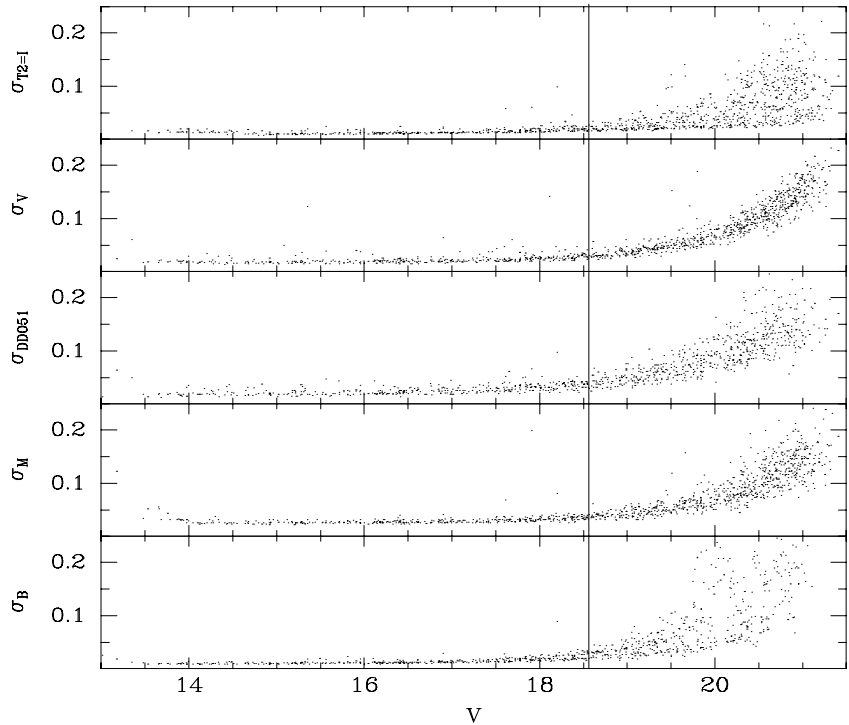}
\caption{Photometric error in all five passbands as a function of $V$ magnitude.  The line indicates
the apparent magnitude of Willman 1's HB, given the properties established in previous literature.}
\end{figure}

Data were reduced with IRAF CCDPROC package and photometered using DAOPHOT (Stetson 1987).  Photometry
was calibrated to the standards of Landolt (1992) and
Geisler (1990, 1996) using iterative techniques described in Siegel et al. (2002).  Both standard and object
stars were measured to total magnitudes using the curve-of-growth analysis of DAOGROW (Stetson 1990).
Figure 2 shows the run of photometric error with $V$ magnitude. The line marks the assumed apparent
magnitude of Willman 1's horizontal branch (HB) assuming $M_V(HB)$ = 0.15 {\rm [Fe/H]} + 0.80 and taking
the M07 [Fe/H] and $(m-M)$ values of -1.5 and 17.9, respectively. The reddening maps of Schlegel et al. (1998) 
indicate the field should have minimal extinction ($E(B-V)=0.01$).  Figure 2 shows that the data
provide good ($\sigma \leq 0.05$) photometry in all passbands for any stars on Willman 1's HB or upper RGB.
Our data do not reach the MSTO of Willman 1, which would be near $V\sim22$.

Accurate stellar positions were measured using the IRAF task TFINDER with the NOMAD astrometric 
catalog (Zacharias et al. 2004).

\section{The Nature of the Willman 1 Overdensity}

\subsection{An Improved Membership Survey}

The abundance distribution and velocity dispersion measured by M07 are key arguments
for Willman 1's classification as a dSph
galaxy and both are inconsistent with the trends established from other dSphs
(Figure 1).  However, surveys of Willman 1 are
hard-pressed to separate out {\it bona fide} 
members of such a low surface-brightness object from the 
strong foreground of Milky Way disk stars. Radial velocities are helpful but
Willman 1's radial velocity is near the
center of the Galactic field star radial velocity distribution (Figure 3), which complicates
membership assignment.  M07 overcome this difficulty by using the equivalent width of Na lines as a
surface gravity discriminant.  However, additional information on stellar surface gravities can
be acquired either directly through high S/N spectra or indirectly 
through a photometric proxy.

\begin{figure}[h]
\epsscale{1.0}
\plotone{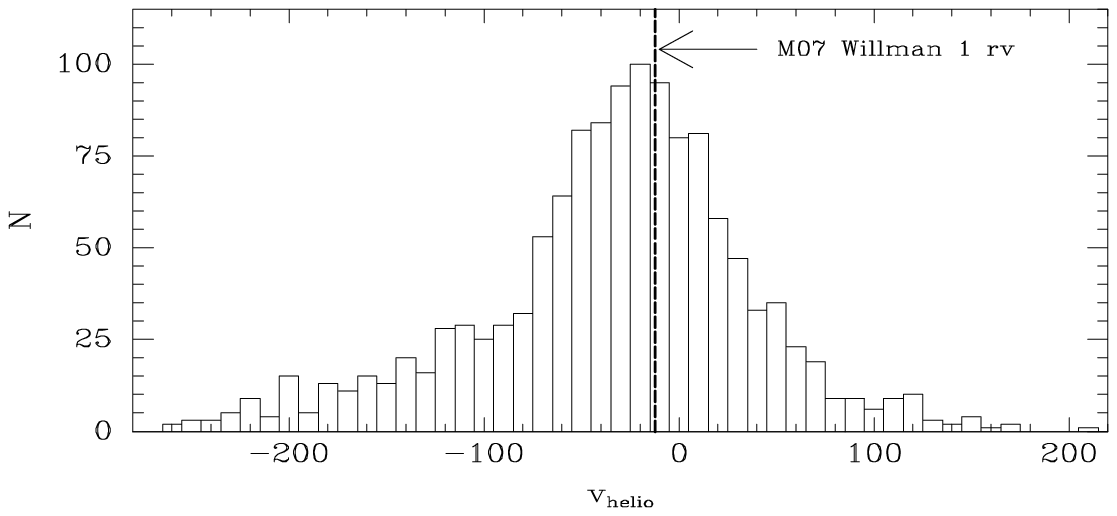}
\caption{The radial velocity distribution of a theoretical Galactic field star populations in the direction
of Willman 1. Galactic simulations were taken from the Besancon model (Robin et al. 2003) and include
the thin disk, thick disk and halo simulated to $V=22$ and out to 50 kpc along the line of sight.  The 
long tail
toward negative velocities is the contribution of the thick disk and halo.  Willman
1's velocity (solid line) is well within the thin disk peak centered at -20 km s$^{-1}$.}
\end{figure}

We used the HET to 
observe two bright stars classified as Willman 1 members by M07 (1269 and 1578 in our Table 1\footnote{
Although the stars we discuss were cataloged by M07, they do not give ID numbers. We refer to these
stars by our own ID numbers.}) in an effort to determine
surface gravities and abundances.  The analysis for these two stars
used the methods and calibrations of Shetrone et al. (2003) but with
the adjustment that the temperature was set to match
photometric temperatures derived using the color-temperature
relations of Ramirez \& Melendez (2005). Input colors were taken from our
own broadband data, the
SDSS DR4 (Adelman-McCarthy et al. 2006) and 2MASS (Skrutskie et al. 2006).

\begin{deluxetable}{ccccccccccc}
\tabletypesize{\scriptsize}
\tablewidth{0 pt}
\tablecaption{Photometric Observations of M07 Spectroscopic Stars}
\tablehead{
\colhead{ID} &
\colhead{$(\alpha,\delta)_{2000.0}$} &
\colhead{$M_0$} &
\colhead{$\sigma$} &
\colhead{$(M-T_2)_0$} &
\colhead{$\sigma$} &
\colhead{$(M-DDO51)_0$} &
\colhead{$\sigma$} &
\colhead{$\Delta_l$} &
\colhead{$P_{RGB}$} &
\colhead{$[Fe/H]_{M07}$}}
\startdata
\multicolumn{11}{c}{M07 Member Stars}\\
\hline
  1137 &  10:48:58.11, +51:02:54.2 &  21.410 &  0.106 &  1.241 &  0.204 &  0.217 &  0.196 &  0.250 & 0.68 &   --\\
  1236 &  10:49:08.10, +51:02:27.4 &  21.100 &  0.156 &  1.387 &  0.189 & -0.003 &  0.208 &  0.107 & 0.50 & -1.1\\
  1269 &  10:49:12.41, +51:05:44.4 &  18.791 &  0.038 &  1.242 &  0.042 & -0.080 &  0.056 & -0.021 & 0.04 & -0.8\\
  1302 &  10:49:15.98, +51:02:26.9 &  20.917 &  0.095 &  1.048 &  0.141 &  0.026 &  0.173 &  0.002 & 0.34 & -1.1\\
  1311 &  10:49:17.44, +51:03:26.1 &  20.532 &  0.050 &  1.005 &  0.076 &  0.128 &  0.082 &  0.080 & 0.42 & -1.7\\
  1334 &  10:49:21.18, +51:03:30.1 &  21.027 &  0.137 &  0.860 &  0.213 &  0.102 &  0.177 &  0.014 & ---  &   --\\
  1386 &  10:49:27.88, +51:03:46.4 &  20.658 &  0.098 &  0.754 &  0.129 &  0.018 &  0.149 & -0.086 & ---  & -1.6\\
  1485 &  10:49:40.85, +51:03:40.4 &  21.283 &  0.174 &  1.177 &  0.240 &  0.248 &  0.217 &  0.253 & 0.69 & -1.4\\
  1578 &  10:49:52.55, +51:03:42.7 &  18.602 &  0.038 &  1.147 &  0.045 &  0.092 &  0.054 &  0.098 & 0.46 & -2.1\\
\hline
\multicolumn{11}{c}{M07 Non-Member Stars}\\
\hline
   971 &  10:48:39.51, +51:04:36.0 &  19.521 &  0.052 &  2.730 &  0.056 &  0.035 &  0.064 &  0.077 & 0.36 &   --\\
  1020 &  10:48:45.14, +51:05:36.4 &  20.657 &  0.070 &  2.484 &  0.078 &  0.062 &  0.123 &  0.190 & 0.70 &   --\\
  1028 &  10:48:46.15, +51:02:12.5 &  18.975 &  0.035 &  2.080 &  0.040 & -0.116 &  0.051 &  0.128 & 0.64 &   --\\
  1074 &  10:48:51.59, +51:03:10.1 &  19.744 &  0.056 &  0.848 &  0.072 &  0.142 &  0.073 &  0.051 & ---  &   --\\
  1091 &  10:48:53.26, +51:06:30.8 &  21.129 &  0.108 &  2.956 &  0.114 &  0.156 &  0.155 &  0.128 & 0.55 &   --\\
  1092 &  10:48:53.66, +51:06:24.1 &  21.510 &  0.098 &  3.468 &  0.102 &  0.253 &  0.161 &  0.286 & ---  &   --\\
  1198 &  10:49:03.89, +51:06:40.4 &  17.366 &  0.026 &  2.487 &  0.029 & -0.086 &  0.040 &  0.051 & 0.15 &   --\\
  1222 &  10:49:06.85, +51:04:23.7 &  18.803 &  0.046 &  0.958 &  0.052 &  0.171 &  0.056 &  0.106 & ---  &   --\\
  1437 &  10:49:33.86, +51:03:33.8 &  20.631 &  0.061 &  1.640 &  0.075 & -0.293 &  0.118 & -0.082 & 0.12 &   --\\
  1496 &  10:49:42.90, +51:04:23.0 &  18.726 &  0.037 &  1.252 &  0.042 & -0.017 &  0.045 &  0.041 & 0.12 &   --\\
  1616 &  10:49:56.99, +51:05:49.6 &  19.055 &  0.040 &  3.113 &  0.043 &  0.155 &  0.055 &  0.109 & ---  &   --\\
  1733 &  10:50:01.32, +51:04:43.5 &  21.168 &  0.086 &  3.530 &  0.090 &  0.170 &  0.167 &  0.245 & ---  &   --\\
  1785 &  10:50:07.88, +51:02:57.6 &  18.869 &  0.038 &  1.547 &  0.043 & -0.163 &  0.051 &  0.016 & 0.09 &   --\\
  1803 &  10:50:09.53, +51:04:15.2 &  19.510 &  0.032 &  2.404 &  0.037 & -0.219 &  0.059 & -0.046 & 0.03 &   --\\
  1853 &  10:50:15.75, +51:02:22.5 &  15.770 &  0.027 &  1.120 &  0.029 & -0.029 &  0.034 & -0.024 & 0.00 &   --\\
\hline
\enddata
\end{deluxetable}

Initial estimates of the surface gravities were made using the distance modulus and
reddening of M07.  As the two stars differ in abundance by 2 dex, we confirmed
each star's derived surface gravity by requiring the abundance determined 
from the ionized species to match the abundance determined
from the neutral species.

For star 1269, we find a radial velocity of $-11.4\pm0.2$, essentially 
identical to the M07 value of $-10.2\pm1.2$. We were able
to determine abundances in star 1269 from 25 Fe\,I lines, 5 Fe\,II lines, 8 Ti\,I lines and
5 Ti\,II lines. When we fit the spectrum using a $logg$=2.5 and the temperature derived 
from NIR colors (5350K), we found
large (3.6$\sigma$) discrepancies between the Fe\,I and Fe\,II abundances for
star 1269.  The optical colors indicated a lower temperature (5000 K) but dropping
the effective temperature that low only marginally improved
the disagreement to 2.4$\sigma$. However, adjusting the surface gravity to that of a main
sequence star ($logg$=4.6) brought the Fe\,I and Fe\,II abundances to within 1 $\sigma$ of each other
at a temperature of 5350K. It is likely
that star 1269 is not a Willman 1 object but is a foreground dwarf. We calculate
a metallicity of [Fe/H]=-0.18 under the assumption that this star is a giant -- a value
significantly more metal-rich than the -0.8 measured by M07.  We also measure
moderate Mg and Ca overabundances 
([Mg/Fe]=+0.21$\pm0.21$, [Ca/Fe]=+0.26$\pm$0.11,
respectively) and a significant overabundance of Ti ([Ti/Fe] = +0.76$\pm$0.21).

For star 1578, we measure a radial velocity of $-20.47\pm1.48$, essentially
identical to the $-22.0\pm0.6$ measured in M07. We
identified 19 Fe\,I lines, 
4 Fe\,II lines, 4 Ti\,I lines and 5 Ti\,II lines in the spectrum. Using
the photometric temperature ($4958\pm95$K in both optical and NIR colors)
and an RGB surface gravity ($logg=2.5$)
produced 1$\sigma$ agreement between the Fe\,I and Fe\,II abundances, confirming that star 1578 is
a low surface-gravity giant. The measured abundances indicate a metallicity of [Fe/H]=$-2.20\pm0.17$ with
$\alpha$-element
abundance ratios slightly below solar ([$\frac{Mg+Ca+Ti}{3Fe}] = -0.11\pm0.09$). This 
$\alpha$-abundance would be 
unusually low for a metal-poor
Milky Way halo star and while the dSph galaxies are known to have low $\alpha$-abundances, -0.11 would
be notably lower than the $\alpha$-abundances of dSph stars 
of comparable [Fe/H] (see, e.g., Figure 2 of
Venn et al. 2004).  This would indicate that the much lower luminosity (and low baryonic mass) of Willman 1
has resulted in much slower chemical evolution
than the Galactic halo or the dSph galaxies. In that scenario, Willman 1 would be unlikely to the remnant of a larger
and therefore faster-evolving system.

Most of M07's stars are too faint for HET-HRS spectroscopy. However, in the absence
of such spectra, the $M$-$T_2$-$DDO51$ photometry system can be used to provide
a proxy measure of surface gravity that has proven useful in identifying sparse populations of RGB
stars against the Galactic foreground (see Majewski et al. 2000a,b; Bessell 2001; Palma et al. 2003; 
Westfall et al. 2006; Sohn et al. 2007;
Siegel et al. 2007).  The three filter system discriminates
high surface-gravity late-type dwarf stars from low surface-gravity, late-type RGB
stars by using the $DDO51$ filter to measure the strength of the gravity-sensitive MgH and Mgb absorption
features (Paltoglou \& Bell 1994). The $M$ filter provides the necessary continuum measure and, in
combination with 
the $T_2=I$ filter, a measure of effective temperature through $M-T_2$ color. The referenced studies
have shown that this three-filter method is more efficient than simple color-magnitude selection
in identifying RGB candidates that are later spectroscopically
confirmed. This is especially true at low surface-brightness levels where
the ratio of contaminants is high.

Figure 4 shows the ($M-T_2$, $M-DDO51$) two-color diagrams for the entire survey and the region
within 5\farcm0 of the nominal Willman 1 photocenter.  Stars have been selected to be PSF-like ($-0.1<
sharp<0.1$; Siegel et al. 2002) or to have small errors ($\sigma_M, \sigma_{T_2}, \sigma_{DDO51} < 0.1$).
The long arc of stars at negative $M-DDO51$ colors
are the field dwarfs and any very metal-rich giants.  Intermediate to metal-poor 
RGB stars normally reside in the space inside this curve, clustered near $(M-DDO51)\sim0$.
However, in contrast to the studies referenced above, there is no prominent RGB in the Willman 1 field.
Rather, there are a few stars near $M-DDO51\sim0$ that represent a combination of Willman 1 
RGB stars, field RGB stars, background galaxies and/or foreground metal-poor dwarf stars.

\begin{figure}[h]
\epsscale{1.0}
\plotone{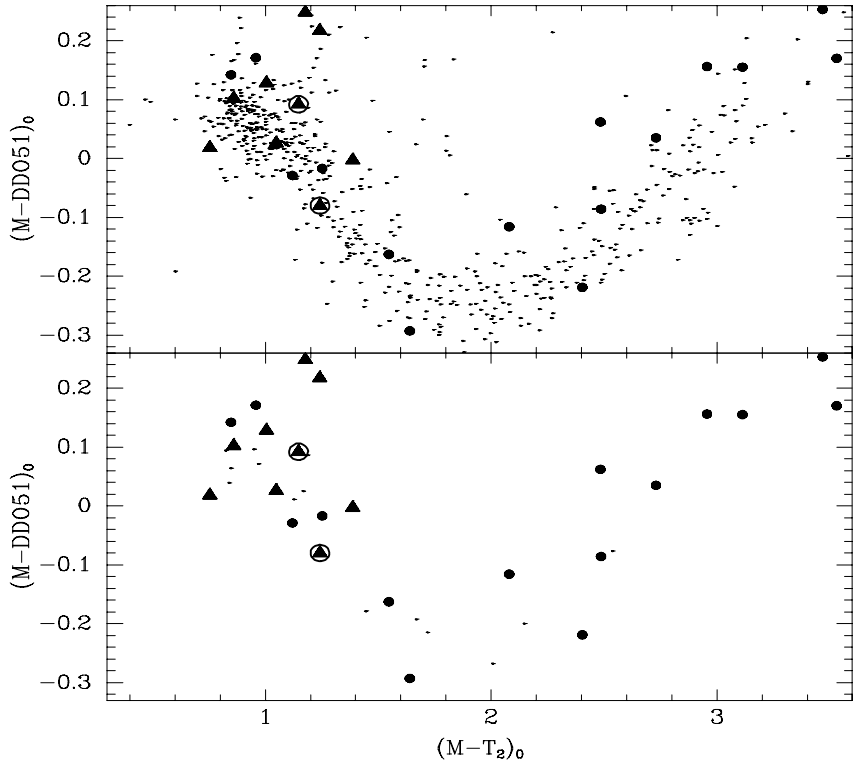}
\caption{Two-color diagrams of the entire Willman 1 field (top) and the region 
within 5' of the nominal photocenter (bottom).  The sample is limited to either stars
with $-0.1 < sharp< 0.1$ and 
$\sigma_M, \sigma_{T_2}, \sigma_{DDO51} < 0.1$ or to stars matched to the M07
catalog.  M07's 
member stars (triangles) and non-member stars (circles) are marked. The two targets observed with
HET-HRS are circled.}
\end{figure}

We have marked on Figure 4 the stars that have counterparts in the M07 survey.
Our photometric
survey matches nine stars M07 classify as members and 15 they classify as non-members.  A preliminary examination
of Figure 4 would indicate that only two of M07's member stars are clearly RGB stars while most of 
the non-members and a significant number of member stars have photometry consistent with dwarfs.

However, this initial impression is deceptive.  Most of M07's stars are quite faint and the photometric
uncertainties for these stars, especially in the critical $DDO51$ filter, are substantial (median 0.14 mag).
We determined the nature of these stars by first
calculating the $\Delta_l$ measure as described in Siegel et al. (2008).  $\Delta_l$ measures
the distance of each star orthogonal to a polynomial fit of the dominant dwarf sequence (Figure 5) over the
color range $1.0 < (M-T_2)_0 < 3.0$.
With the problem reduced to a single dimension, we calculate how much of each star's Gaussian
probability distribution function lies within the RGB box.\footnote{No color-magnitude selection was applied
owing to the sparseness of the Willman 1 RGB.}

\begin{figure}[h]
\epsscale{1.0}
\plotone{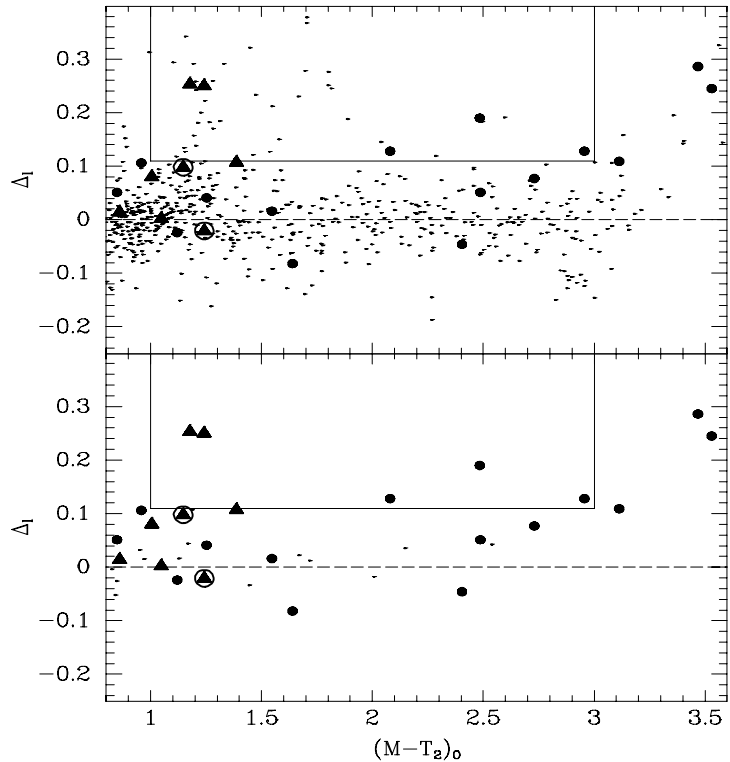}
\caption{$\Delta_l$ diagrams that measure the photometric distance of each star from a fitted dwarf
locus over the color range $1.0 < (M-T_2)_0) < 3.0$.  The sample is limited to either stars with $-0.1 < sharp< 0.1$ 
and $\sigma_M, \sigma_{T_2}, \sigma_{DDO51} < 0.1$ or to stars matched to the M07
catalog.  M07's 
member stars (triangles) and non-member stars (circles) are marked. The two targets observed with
HET-HRS are circled. The box marks our RGB selection criteria.}
\end{figure}

Table 1 lists magnitude, errors, $\Delta_l$ measures and the likelihood that each M07 star
in our program
has true photometric properties within the RGB locus.  Many of M07's stars are close to 50\%,
indicating that our photometry of these stars is too poor for reliable classification.  Additionally, a 
number are too blue, being close to the MSTO where the Mg absorption features vanish.
However, we find that seven of M07's member stars are within the $(M-T_2)_0$ color range
at which Mg features manifest.  Two of these stars have less than a 35\% chance of residing
within the RGB color-color locus, including star 1269, confirming the spectral analysis.
Three are ambiguous, with probabilities between 42\% and 50\%, including star 1578. Two
are most likely RGB stars with probabilities greater than 68\%.

Of M07's non-member stars, ten are in the usable color range.  Six of these stars are clearly
dwarfs ($\leq$ 15\%), two are ambiguous (36\% and 55\%) and two are clearly within the RGB locus ($>$ 64\%)
-- although well-removed from Willman 1's RGB (Figure 6).

We used our classification methods to identify any RGB stars missed in the M07 study and
found three bright RGB candidates ($\Delta_l > .105; P_{RGB}>0.7; M<19$). All three 
are well outside Willman 1's $r_h$ but lie near the hypothetical Willman 1 RGB.  The properties of these stars, 
including the rvs measured with HET-HRS, are listed in Table 2.  Radial velocity uncertainties 
are $\sim$1 km s$^{-1}$.  As can be seen, none of these three stars lies at the nominal Willman 1 
radial velocity. They
resemble a typical high-dispersion halo population, meaning they are either foreground subdwarfs or
unrelated halo RGB stars.

\begin{deluxetable}{ccccccccccc}
\tabletypesize{\scriptsize}
\tablewidth{0 pt}
\tablecaption{HET Radial Velocities of Washington-Selected Stars}
\tablehead{
\colhead{ID} &
\colhead{$(\alpha,\delta)_{2000.0}$} &
\colhead{$V_0$} &
\colhead{$(B-V)_0$} &
\colhead{$(V-I)_0$} &
\colhead{$M_0$} &
\colhead{$(M-T_2)_0$} &
\colhead{$(M-DDO51)_0$} &
\colhead{$\Delta_l$} &
\colhead{$P_{RGB}$} &
\colhead{$v_{hel}$}}
\startdata
 385  & 10:47:26.53, +51:05:06.8  & 15.194 & 0.726 & 0.901 & 15.494  & 1.202  & 0.112 & 0.138 & 0.79 & -74.5\\
 849  & 10:48:23.16, +51:19:20.5  & 16.954 & 0.819 & 0.964 & 17.269  & 1.280  & 0.112 & 0.170 & 0.93 & -124.0\\
1304  & 10:49:16.56, +51:14:20.9  & 17.651 & 0.779 & 0.933 & 17.981  & 1.263  & 0.133 & 0.182 & 0.94 & +53.7\\
\enddata
\end{deluxetable}

Figure 6 shows the $VI$ color-magnitude diagram of the Willman 1 field. The bright stars are
taken from the photometric survey described in \S2.2 while the faint stars are taken from M07's INT-WFC
study.  The likely HB magnitude is indicated by the 
dotted line while theoretical isochrones from Dotter et al. (2007) set 
at [Fe/H]=-1.5 and -2.2, [$\alpha$/Fe]=+0.0,
14 Gyr are overlayed for comparison. Large points indicate RGB likelihood. M07 member 
stars are marked as triangles;
non-members as circles. Filled symbols are those we identify as RGB stars; half-filled are those  either
outside the useable range of $(M-T_2)_0$ color or with uncertain classification; empty symbols are those we identify
as non-RGB stars. The starred points mark the objects identified as potential giants in Table 2 that
have radial velocities inconsistent with Willman 1 membership.
We have also marked with a filled triangle the HB star
noted by W05 (SDSSJ104913.15+5210232.6).  SDSS DR6 indicates this star has a radial velocity 
of $\sim$-20 km s$^{1}$, making it a likely Willman 1 member (see \S3.3).

\begin{figure}[h!]
\epsscale{1.0}
\plotone{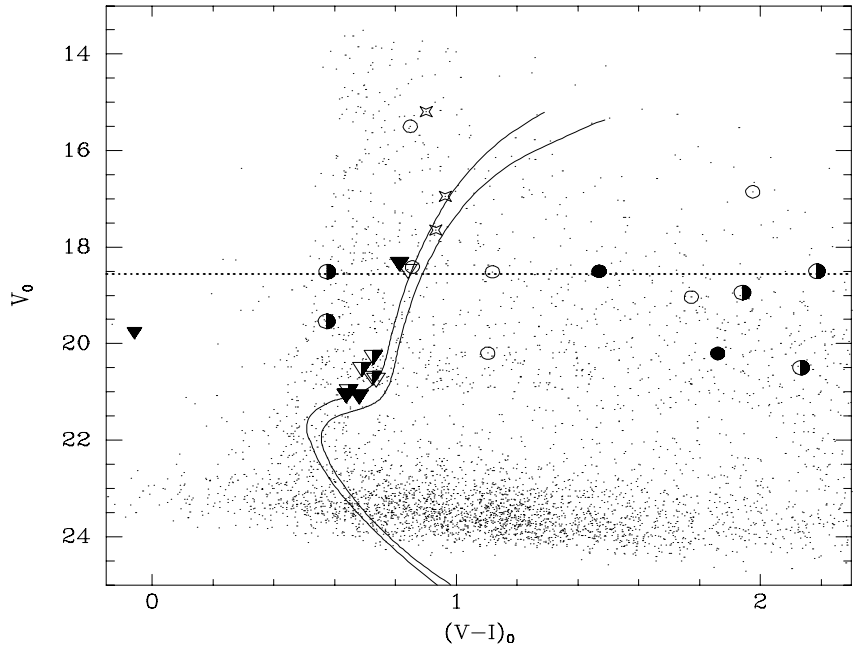}
\caption{$VI$ color-magnitude diagram of stars surveyed by M07 and this study. Triangles are stars
identified as Willman 1 members in M07; circles are those identified as non-members. Empty symbols
are those we identify as non-giants; half-filled symbols are ambiguous; filled symbols 
are those we identify as giants. Star symbols are Washington-selected RGB candidates observed with
HET, all three of which appear to be non-members.  The solid
lines are theoretical isochrones from Dotter et al. (2007) while the dotted line represents
the likely magnitude of the HB. The solid triangle near the HB is the member star identified in
W05.}
\end{figure}

As can be seen, a significant number of M07's stars, including one of their brightest
members, disappear from the color-magnitude diagram when the $M$-$T_2$-$DDO51$ photometric and
HET-HRS information are included. No new members are added by the 0.6 square degree photometric survey.
The confirmed member stars lie almost exclusively on the metal-poor isochrone -- although
a younger metal-rich isochrone would also fit the data.

\subsection{A Search for RR Lyrae Variable Stars}

RR Lyrae variable stars can provide both a precise distance measure to and age/abundance information
about their parent stellar population (see, e.g., Siegel 2006). Additionally, RR Lyrae
stars can be used as tracers to explore the extended structure of dSph galaxies
(see, e.g., Kuhn et al. 1996).

In an attempt to identify variable stars in the Willman 1 field, we obtained ten $B$-band
epochs over the course of the two MDO 0.8m observing runs.  A simple plot of 
the ratio of measured dispersion over formal photometric error (Figure 7) shows that there is one highly 
variable star (at a significance of 68$\sigma$)
in the Willman 1 field with an approximate amplitude of $A_B\sim1.4$.  All other stars 
vary within 1-2 $\sigma$ of their mean values.  Of the three 
remaining stars with variability
greater than twice their formal error, two are bright ($B\sim14.5$) and are either saturated or have 
underestimated formal errors.  The remaining star has a single discrepant measure.
We ran a similar analysis of
the $DDO51$ data, which consists of ten epochs summed to produced the deep $DDO51$ image
for our photometric analysis (\S3.1). While these data were obtained in a shorter span
of time (two nights) and are therefore less optimized for variable-star searching, they show
a similar pattern (lower panel of Figure 7) to the $B$ data and identify the same variable
star.

\begin{figure}[h]
\epsscale{1.0}
\plotone{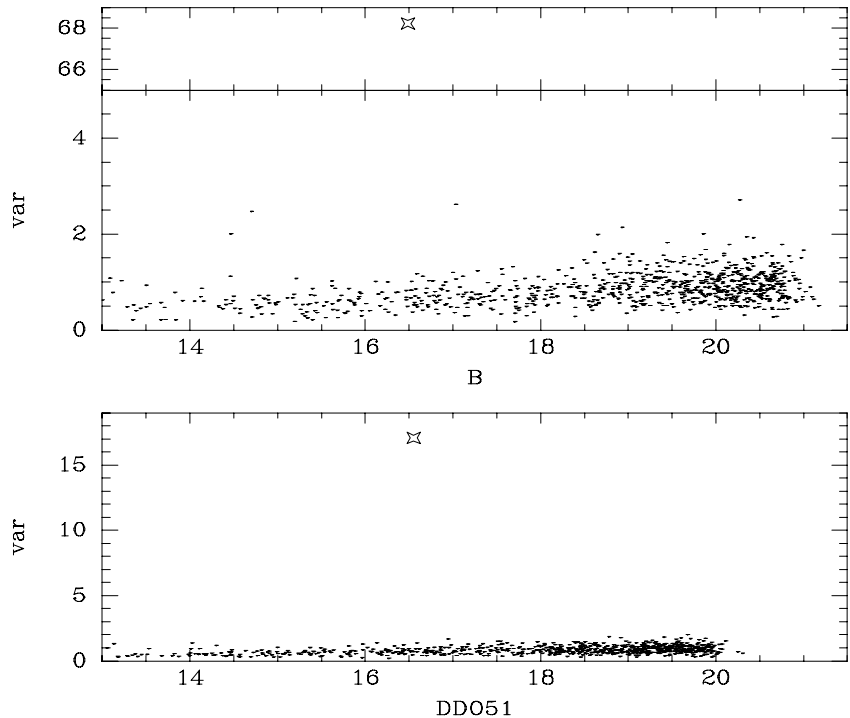}
\caption{Variability of stars within the Willman 1 field.  The starred point indicates the variable star.  No
other object has a variability greater than three times its formal error.}
\end{figure}

We attempted to fit the light curve using the fitting
code and templates of Layden (1998) and Siegel (2006) but found the fit periods to be degenerate.
While the data are not extensive enough to break the period degeneracy, they indicate
that this variable is unlikely to be associated with Willman 1.
First, the variable star has a mean magnitude of $<B>=17.02$, two 
mag brighter than the likely HB of Willman 1 (see Figure 8).
Second, the variable is over 20' from the nominal Willman 1 center -- ten
times the $r_h$ measured by W05.  Although located within $30^{\circ}$ of the apparent Willman 1 major axis
(see Figure 3 of W07), is it unlikely that the only variable star in Willman 1 would be so far
removed from the photocenter.

The simplest and most likely explanation is that the variable star is a foreground RR Lyrae star.  It is 
possible, but less likely, that it is an anomalous cepheid (AC, Pritzl et al. 2004) associated with Willman 1
and embedded in an extended stellar distribution.  The specific frequency 
of ACs has been shown to increase with declining
mass and abundance (Mateo et al. 1995; Pritzl et al. 2004). However, Willman 1 is so small that even if
the Pritzl et al. absolute magnitude-specific frequency relation were extrapolated
over another {\it six} magnitudes, it would predict less than one AC. The abundance-specific 
frequency relation would predict zero ACs, whether evaluated at [Fe/H]=-1.5 or -2.2.

\subsection{The Extended Structure of Willman 1}

Figure 8 shows the $BV$, $VI$ and $MT_2$ color-magnitude diagrams of the Willman 1 field.  These diagrams 
have been cleaned 
to retain only objects with DAOPHOT {\it sharp} parameters between -0.1 and 0.1 and photometric
errors below 0.15 mag.
The HB magnitude and isochrones are identical to those in Figure 6.
The starred point indicates the variable star identified in \S3.2 while the squares indicate the
BHB star identified in W05 from SDSS and the RGB star we confirm in \S3.1.
Consistent with previous investigations and \S3.1, we find little evidence of a prominent
RGB in the 0.6 square degree PFC field.  There is a slight overdensity where the lower part of the
Willman 1 RGB is located.
However, these faint blue objects show a flat spatial distribution across the field and 
are therefore more likely to represent the metal-poor turnoff of the halo field stars or compact background
galaxies.

\begin{figure}[h]
\epsscale{1.0}
\plotone{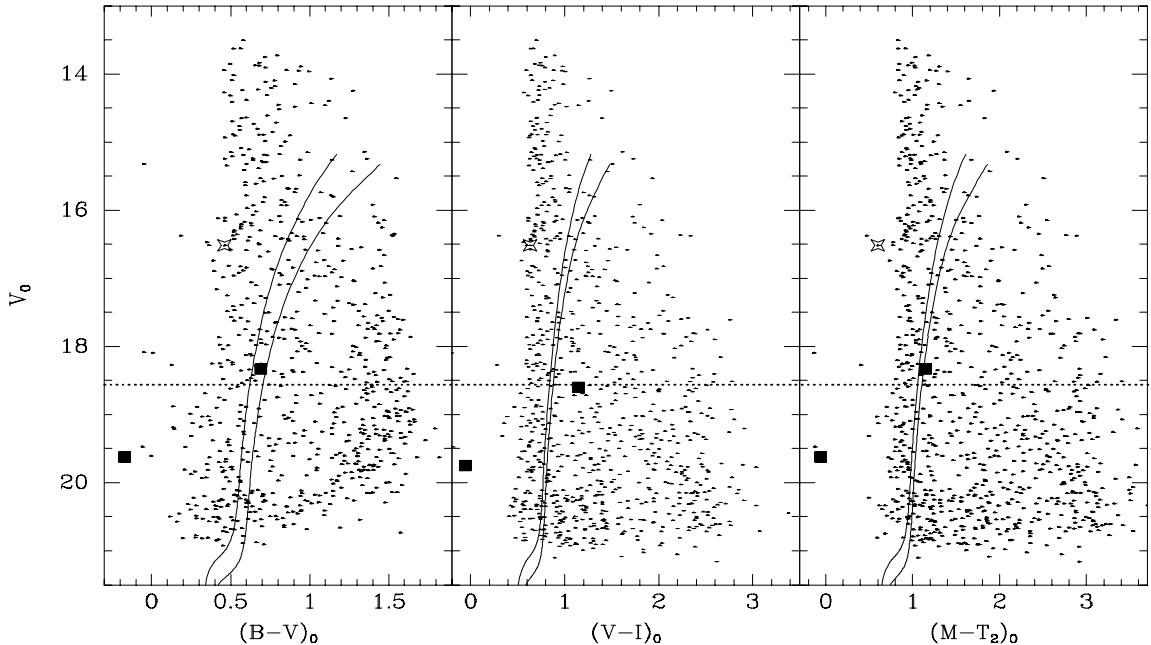}
\caption{$BV$, $VI$ and $MT_2$ color-magnitude diagrams of the Willman 1 field.  Star are selected
to have $-0.1 < sharp < 0.1$ and $\sigma_V,\sigma_{T_2},\sigma_M <0.15$.
The dotted line indicates
the hypothetical magnitude of the HB. The starred point is the variable star identified in \S3.2; the squared points
the BHB star identified in W05 and the RGB star identified in \S3.1.  The solid
lines are theoretical isochrones from Dotter et al. (2007). Note
the lack of any apparent RGB or BHB.}
\end{figure}

The CMDs show about a dozen stars slightly bluer than the field star edge
with $V-I$, $B-V$ and $M-T_2$ colors similar to BHB stars. A handful of these
are clumped at the faint end of the data near V$\sim19.4$ and could potentially be a diffuse Willman 1 HB
if the distance modulus has been badly underestimated. However, the number and magnitude distribution
of these objects would also be consistent with the number
and magnitude distribution of faint blue field stars we have seen
in similar surveys of other dSph galaxies such as Bootes (Siegel 2006). They also do not show
any central concentration in the field.

That these faint blue objects are not Willman 1 HB stars is demonstrated in Figure 9, which
compares the raw $VI$ field CMD to one overlayed with both isochrones and synthetic
horizontal branches (SHB) from Dotter et al. (2007). The two SHBs have 50 stars
each and are set at [Fe/H]=-1.5 and -2.2,
[$\alpha$/Fe]=+0.0. HB mass is set at the mass of the TRGB for the relevant isochrone with an average 
mass loss of $0.05\pm0.05$ $M_\odot$.

\begin{figure}[h]
\epsscale{1.0}
\plotone{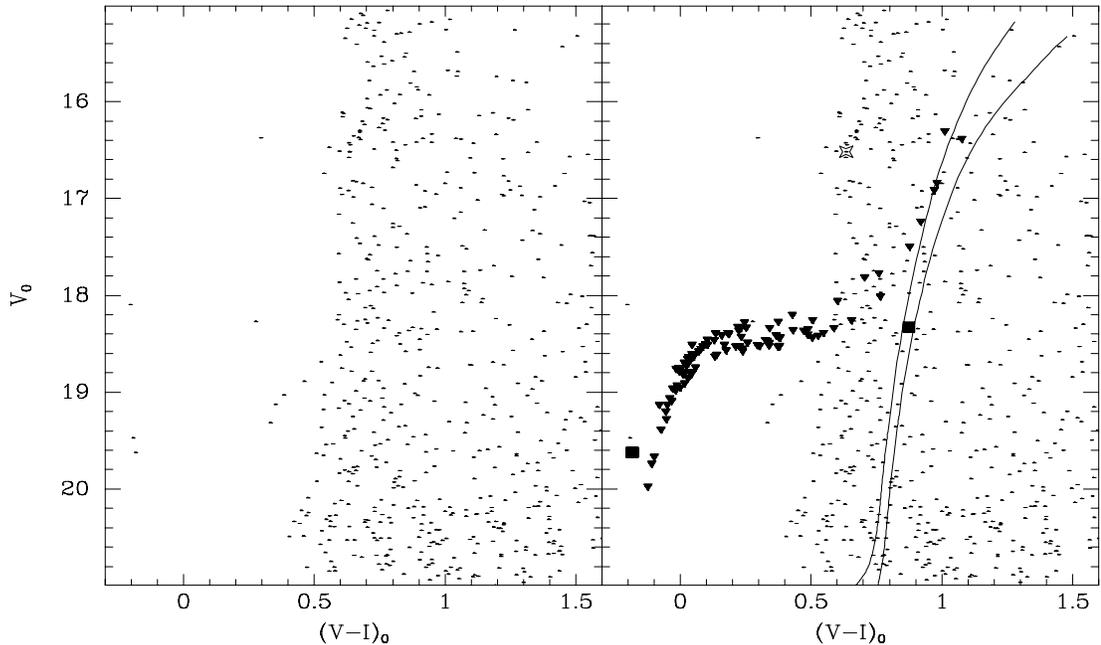}
\caption{The raw $VI$ color-magnitude diagram of the Willman 1 field (left) and the same CMD overlayed
with isochrones and synthetic horizontal branches from Dotter et al. (2007) corresponding to the 
possible Willman 1 populations. Symbols are as in Figure 8.}
\end{figure}

The apparent sequence of faint blue stars is well below the magnitude of the SHB and has a downward slope
that is not seen in the synthetic model.  In fact, only three objects are within 0.5 
magnitudes of the SHB.  The BHB star
identified in W05 lies slightly bluer and brighter than the SHB. As noted in \S3.1, this star
is at the Willman 1 radial velocity.
Its location above the SHB could indicate either that Willman 1 is closer
than 40 kpc or that the star is evolving away from the BHB. A second star
is within 7' the Willman 1 photocenter and along the metal-poor SHB but has a $V-I$ color
of 0.28. Our studies of other dSph galaxies (see, e.g., Siegel 2006) have shown this color to lie near or in
the instability
strip.  Since this star does not vary, it is unlikely it is a Willman 1 HB
star.  Finally, a third star, located more than 14' from the Willman 1 photocenter, is slightly brighter 
and bluer than the W05 HB star.  While this star could be a Willman 1 member, its membership would require
that the star be both extratidal {\it and} unusually blue and evolved. It
is highly unlikely that an object as faint as Willman 1 has two such rare BHB stars.
In short, it is likely that Willman 1 does not have any BHB stars 
within the PFC field beyond the one identified in W05. Willman 1 could, however,
have non-variable RHB stars lost within the field star blue edge.

When combined with the negative detection of extratidal giants (\S3.1) or RR Lyrae (\S3.2), this indicates that
our thorough 0.6 square degree survey identifies no measurable extended structure in Willman 1. Not a single
confirmed RGB, RR Lyrae or BHB star lies beyond the measured optical limit.
The only confirmed members are the W05 HB star and some of the M05 giants, all of which 
are within 5' of the Willman 1 photocenter.


\section{Discussion and Conclusions}

The paucity of bright spectroscopic targets in Willman 1 and the difficulty in using radial 
velocities alone to separate
out members makes conclusions about its nature and properties tentative. However, our survey
results
have several important implications about the nature of Willman 1.

$\bullet$ The metallicity spread implied by M07 is now less certain. The most metal-rich M07 star, and
the only metal-rich member with $S/N>13$ in M07, is clearly a foreground dwarf star.  Another 
metal-rich M07 star, 1302, is close to the dwarf locus, with an RGB probability of 34\%.
Without star 1269, the metallicity
dispersion of Willman 1 drops to 0.3 dex, which is only slightly greater than the measurement error 
(estimated by M07 to be 0.2 dex at S/N=15). Moreover, the remaining metal-rich Willman 1 members 
are close to the MSTO where the CaT abundance method used by M07 it not very reliable since it is designed
for use with late-type RGB stars (Rutledge et al. 1997; Carretta \& Gratton 1997). In fact,
the only confirmed member of Willman 1 with an M07 $S/N>15$ has an 
abundance of [Fe/H]=-2.1, which we confirm directly from the Fe lines (as [Fe/H]=-2.2). {\it The existing
evidence does not
support an abundance spread in Willman 1 and hints at a low, rather than intermediate abundance}.
Figure 10 shows the revised $M_V-[Fe/H]$ trend of the dSph galaxies, assuming a -2.2 metallicity for Willman 1
and shows that a metal-poor Willman 1 would be consistent with the trend established
in other very low-luminosity dSph galaxies (Munoz et al. 2006; M07, SG07).  The flattening of the
luminosity-metallicy relation -- and many of the brighter dSph galaxies with multiple
populations having similar trace metal-poor populations -- could support the hypothesis of 
a common origin for many
of the dSph galaxies (Kunkel \& Demers 1977; Lynden-Bell 1982; Majewski 1994; Lynden-Bell \& Lynden-Bell
1995; Metz \& Kroupa 2007; Metz et al. 2007).  Alternatively, this could be a threshold
metallicity for the formation of stellar populations.

$\bullet$ Our study does not affect the inferred velocity dispersion of Willman 1.  The nine member 
stars we match to 
our Washington photometry have a velocity dispersion of $4.4\pm1.3$ km s$^{-1}$.  When star 1269 is removed, the remaining
M07 stars have a dispersion of $4.5\pm1.5$ km s$^{-1}$.  Further
pruning of the sample -- even down to the extreme of the three stars that are either spectroscopically or 
photometrically classified as giants -- does not reduce the velocity dispersion.

\begin{figure}[h]
\epsscale{1.0}
\plotone{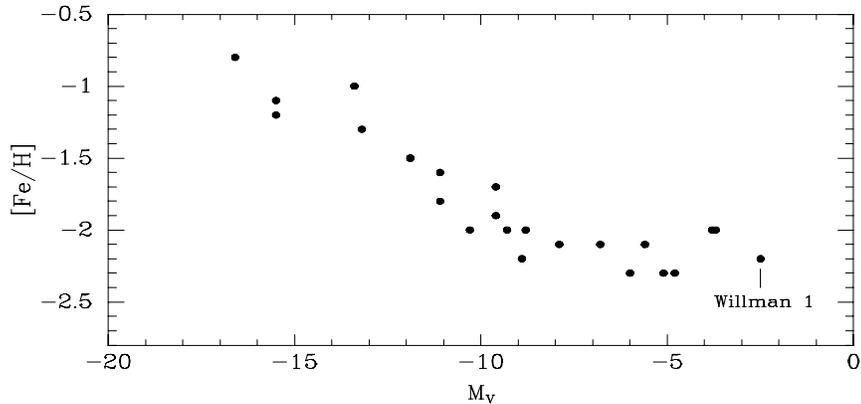}
\caption{The revised $M_V$-$[Fe/H]$ relation for dSph galaxies, assuming star 1578 indicates the
true abundance of Willman 1. Contrasted against Figure 1, Willman 1's abundance is now consistent
with dSph galaxies of similar luminosity.}
\end{figure}

$\bullet$. We measure a slightly sub-solar $\alpha$-abundance in star 1578, a value similar to but lower 
than the $\alpha$-abundances measured in stars of similar [Fe/H] in other dSph galaxies (Venn et al. 2004)
and globular clusters thought to have been tidally stripped
from dwarf galaxies (e.g., Palomar 12; Dinescu et al. 2000; Cohen 2004). This indicates that Willman 1 has
a chemical evolution history similar, but not identical, to that of other dwarf galaxies. It certainly
formed in a different environment than the Milky Way field stars or globular clusters. A slow chemical
evolution would be consistent with Willman 1 having always been a low-mass object.

$\bullet$ While the isophotes of W07 appear to show an irregular object, the 
wide-field 0.6 square degree survey of Willman 1 fails to conclusively identify a single bright RGB star, 
a single RR Lyrae star or a single BHB star outside of the half-light radius ($1\farcm9$).
In fact, we identify {\it no} Willman 1 members beyond those identified in M07 (several of which
appear to be non-members). The lack of extratidal stars supports the structural analysis
of M07 and W07, which, using the more numerous stars
of the upper main sequence, show 
radial profiles that are roughly consistent with a King (1962) or exponential profile with no indication
of extended structure.

$\bullet$ It is important, however, to realize that the extant data are limited in their ability to probe the
extended structure of Willman 1.
W07 and M07 only trace Willman 1 over two orders of magnitude and barely reach the foreground starcount 
level in the outer isophotes. Such
surveys can lack the sensitivity needed to trace low surface-brightness extended structure 
(see, e.g., Siegel et al. 2008). While tracer studies of evolved stars are more
sensitive to low surface-brightness extensions, Willman 1 is so small and sparse that
even the $DDO51$ photometry provides no additional information on the surface brightness profile. Only a deeper
survey using an 8-m class imager would be able to trace Willman 1's outer isophotes to fainter
levels using faint main sequence stars.

These properties are consistent with a low-mass dark matter-dominated dSph galaxy. However, it is worth
noting that they would also
be consistent with a low-mass globular cluster in the late stages of tidal disruption by the Milky Way.
Willman 1 bears some resemblance to
the globular cluster Palomar 13. Pal 13, like Willman 1, is small, diffuse and shows
signs of being tidally disrupted (Siegel et al. 2001). Cote et al. (2002) measured
an apparent M/L ratio of 40 for Pal 13.  While this can be interpreted as an indication of dark
matter, objects in the final throes of tidal disruption can also have an inflated
core velocity dispersion (Munoz et al. 2007).  Moreover, Blecha et al. (2004)
and Clark et al. (2004) have shown that Palomar 13 has a high binary
fraction, possibly resulting from the preferential loss of low-mass stars, which inflates the measured
velocity dispersion. This is an important point to keep in mind with Willman 1, which, like
Palomar 13, could have an inflated velocity dispersion owing to its overlap with the disk radial velocity
distribution, its sparseness and/or a high binary fraction.  A binary
fraction enhanced by mass loss would also account for the broad MS observed in M07 and W07, obviating the 
need for an abundance spread.

Determining if Willman 1 is a low-luminosity dSph or a disrupting globular cluster remains just
out of reach of the present studies but is amenable to further investigation.
Deep spectroscopy from next-generation telescopes would allow 
abundances to be measured for the MSTO
and MS stars, minimizing the problem of foreground dwarf contamination.  Further spectroscopy and more
extensive photometry would reveal if Willman 1, like Palomar 13, has a high binary fraction that is inflating
its velocity dispersion.  Proper motions would
allow separation of member stars from the field and determine the orbit of Willman 1, revealing
much about its dynamical past.

As more low-luminosity objects are discovered in the SDSS, the parameter-space defined by these objects will
become clearer.  For the moment, Willman 1 still remains in the nebulous parameter space between dSph 
galaxies and globular clusters. It has yet to definitively reveal its nature.

\acknowledgments

Support for this work was provided by NSF grant AST-0306884
We would like to thank Steve Odewahn, Eusebio Terrazas, John Caldwell, 
Heinz Edelmann, Vicki Riley and Frank Deglman for their assistance in 
collecting the HET spectra.
The Hobby-Eberly Telescope (HET) is a joint project of the 
University of Texas at Austin, the Pennsylvania State University, Stanford University, Ludwig-Maximilians-
Universit\"{a}t M\"{u}nchen, and Georg-August-Universit\"{a}t G\"{o}ttingen.
The HET is named in honor of its principal benefactors, William P. Hobby and Robert E. Eberly.

\end{document}